\documentclass[aps,showpacs,twocolumn,pra,superscriptaddress]{revtex4-1}
\usepackage[english]{babel}
\usepackage[latin1]{inputenc}
\usepackage{hyperref}
\usepackage{amsmath, amssymb, amsfonts}
\usepackage{amssymb,xcolor}
\usepackage{graphicx}
\usepackage{multirow}
\graphicspath{{images/}}
\usepackage{exscale}
\usepackage{graphicx}
\usepackage{latexsym}
\newcommand{\ket}[1]{|#1\rangle}
\newcommand{\bra}[1]{\langle #1|}
\newcommand{\Tr}{\mathrm{Tr}}

\newcommand{\n}{\nonumber\\}

\begin{document}
\title{The geometric structure of quantum resources for Bell-diagonal states}
\author{Jin-Xing Hou}\email{jinxhou@163.com}
\affiliation{Institute of Modern Physics, Northwest University, Xi'an 710069,  China}
\affiliation{Shaanxi Key Laboratory for Theoretical Physics Frontiers, Xi'an 710069, China}
\affiliation{School of Physics, Northwest University, Xi'an 710069, China}
\author{Yun-Long Su}
\affiliation{Institute of Modern Physics, Northwest University, Xi'an 710069,  China}
\affiliation{Shaanxi Key Laboratory for Theoretical Physics Frontiers, Xi'an 710069, China}
\affiliation{School of Physics, Northwest University, Xi'an 710069, China}
\author{Si-Yuan Liu}
\affiliation{Institute of Modern Physics, Northwest University, Xi'an 710069,  China}
\affiliation{Shaanxi Key Laboratory for Theoretical Physics Frontiers, Xi'an 710069, China}
\author{Xiao-Hui Wang}
\affiliation{Shaanxi Key Laboratory for Theoretical Physics Frontiers, Xi'an 710069, China}
\affiliation{School of Physics, Northwest University, Xi'an 710069, China}
\author{Wen-Li Yang}
\affiliation{Institute of Modern Physics, Northwest University, Xi'an 710069,  China}
\affiliation{Shaanxi Key Laboratory for Theoretical Physics Frontiers, Xi'an 710069, China}

\begin{abstract}
Two-qubit Bell-diagonal states can be depicted as a tetrahedron in three dimensions.
We investigate the geometric structure of quantum resources, including coherence and quantum discord, in the tetrahedron.
The ordering of different resources measures is a common problem in resource theories, and
which measure should be chosen to investigate the structure of resources is still an open question.
We consider the geometric structure of quantum resources
which is not affected by the choice of measure.
Our work provides a complete structure of coherence and quantum discord for Bell-diagonal states.
The pictorial approach also indicates how to explore the structure of resources
even when we don't have consistent measure of a concrete quantum resource.

\end{abstract}

\maketitle
\section{Introduction}
Quantum correlations
including entanglement, discord and coherence have been widely studied over the last three decades.
In the early days of quantum information theory, entanglement \cite{R. Horodecki}
was regarded as a basic resource that can achieves many tasks which are impossible within
the framework of classical physics, such as
quantum key distribution \cite{C. H. Bennett}, teleportation \cite{C. H. Bennett2},
and superdense coding \cite{C. H. Bennett3}.
However, entanglement does not capture all quantum characteristic
since some separable states still contain quantum resources, i.e.,
they are not entirely classical \cite{H. Ollivier}.
And later another type quantum resource
beyond entanglement called quantum discord \cite{A. Streltsovdiscord},
which has been proven to be an important quantum resource
in quantum-information processing tasks \cite{M. Piani,A. Ferraro,F. Altintas,Y. Yeo,B. Bellomo,Y. Li},
was proposed.
Recently,
quantum coherence,
which marks the departure of quantum theory from classical physics,
is considered to be a equally
important resource in quantum physic.
The resource theory of coherence has been widely studied \cite{T. Baumgratz,A. Winter,Fanheng1,A. Streltsov,shi}
since the quantification of coherence\cite{T. Baumgratz}.

Two-qubit Bell-diagonal states, depicted as a tetrahedron in three dimensions,
are significant for understanding states with more complex structure.
Resource theories for Bell-diagonal states have attracted many attentions
in recently years \cite{Horodecki,T. R. Bromley,L. Mazzola}.
The geometric structures of entanglement and discord for Bell-diagonal states
have been depicted by the level surface of
concurrence and quantum discord nonanalytic \cite{Wootters,Matthias D. Lang}.
The purpose of this paper is to investigate
the structure of coherence and quantum discord
for Bell-diagonal states in the tetrahedron explicitly.

Based on the rigorous framework of coherence proposed by Baumgratz \emph{et al}. \cite{T. Baumgratz},
several reasonable measures \cite{D. Girolami,A. Streltsov1,L.-H. Shao,X. Yuan} have been put forward.
In this paper, we first consider two
well-known coherence measures, namely, the $l_1$ norm of coherence
and relative entropy of coherence, which are simple and
useful tools to uncover various characteristics of
quantum coherence \cite{U. Singh,S. Cheng,M. N. Bera,A. Mani}.
We show that the relative entropy of coherence and $l_1$ norm of coherence
do not give the same ordering for all Bell-diagonal states.
Which measure should be used to quantify the coherence of
Bell-diagonal states is a puzzling problem.
To avoid this problem, we divide the tetrahedron into countless rays
which are the trajectory of Bell diagonal
states under incoherent quantum channels,
and the coherence of the states in those rays are not frozen.
Monotonicity of coherence insures that the states limited in any one of those rays have the same ordering for all coherence measures.
Moreover, quantum discord\cite{H. Ollivier} and geometric quantum discord\cite{Daki} give
the same ordering for the states limited in any one ray towards the center of tetrahedron.

We indicate the complete structures of coherence and quantum discord, respectively,
which are identical for different measures of coherence and quantum discord.
Moreover, we analyse the properties of level surface of coherence and discord in Bell-diagonal states.
The explicit structure of quantum resources
can be used as the guidance for studying the behavior of resources under quantum channels,
such as the frozen of resources \cite{T. R. Bromley}, sudden death \cite{A. Ferraro}
and sudden transition \cite{L. Mazzola}.

This paper is arranged as follows.
In Sec. \ref{aaa},
we investigate the ordering of coherence for Bell-diagonal states.
In Sec. \ref{bbb},
we consider the evolution of states under incoherent quantum channels and obtain the structure of coherence in three-parameter space.
In Sec. \ref{ccc},
we research the structure of discord for Bell-diagonal states by means of the ordering-preserving states.
Finally, we summarize our results in Sec. \ref{eee}.
\section{the ordering of coherence for Bell diagonal states}\label{aaa}

\begin{figure}{}
  \centering
  \includegraphics[width=6cm,height=6cm]{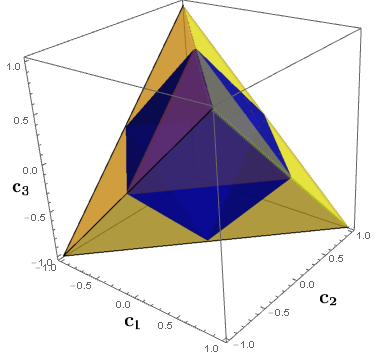}\\
  \caption{Two-qubit Bell-diagonal states described by parameters $c_1,c_2,c_3$ can be depicted as a tetrahedron $\mathcal{T}$. The blue octahedron $\mathcal{O}$ in tetrahedron, specified by $|c_1|+|c_2|+|c_3|\leqslant1$, is the set of separable Bell-diagonal states. There are four
entangled regions outside $\mathcal{O}$, in
each of entangled region the biggest eigenvalue $\lambda_{ab}$ is the one associated
with the Bell state at the vertex. Zero discord states lie on the axes and incoherent states lie on $c_3$ axis.}
\label{sep}
\end{figure}

The Bell-diagonal states of two qubits with
a computational base \{$\ket{00},\ket{01},\ket{10},\ket{11}$\},
have the form \cite{Bell state,Horodecki}
\begin{equation}\label{}
  \rho_{AB}=\frac{1}{4}(I\otimes I+\sum_{j=1}^3c_j\sigma_j\otimes\sigma_j)=\sum_{ab}\lambda_{ab}\ket{\beta_{ab}}\bra{\beta_{ab}},
\end{equation}
with the corresponding density matrix of $\rho_{AB}$ to be
\begin{equation}\label{}
  \rho_{AB}=\frac{1}{4}\left(
    \begin{array}{cccc}
      1+c_3 & 0 & 0 & c_1-c_2 \\
      0 & 1-c_3 & c_1+c_2 & 0 \\
      0 & c_1+c_2 & 1-c_3 & 0 \\
      c_1-c_2 & 0 & 0 & 1+c_3 \\
    \end{array}
  \right),
\end{equation}
where \textit{I} is the identity operator on the subsystem and the matrices $\sigma_{j}$ are the Pauli spin matrices and $c_{j}$ are real numbers. The eigenstates of $\rho_{AB}$ are the
four Bell states \cite{Matthias D. Lang}
\begin{equation}\label{}
  \ket{\beta_{ab}}\equiv(\ket{0,b}+(-1)^a\ket{1,1\oplus b})/\sqrt{2}
\end{equation}
 with eigenvalues
\begin{equation}\label{}
  \lambda_{ab}=\frac{1}{4}[1+(-1)^ac_1-(-1)^{a+b}c_2+(-1)^bc_3],
\end{equation}
where $a\in\{0,1\},b\in\{0,1\}$.
The real numbers $c_j$ are limited
in a tetrahedron $\mathcal{T}$, as showed in Fig. \ref{sep}.

For two measures of coherence $\mathcal{C}_A(\rho)$ and $\mathcal{C}_B(\rho)$, if the following condition
\begin{equation}\label{order}
  \mathcal{C}_A(\rho_1)\leq(\geq)\mathcal{C}_A(\rho_2) \Leftrightarrow \mathcal{C}_B(\rho_1)\leq(\geq)\mathcal{C}_B(\rho_2),
\end{equation}
is satisfied for any arbitrary states $\rho_1$ and $\rho_2$, those two measures give the same ordering, otherwise they do not.
The following method can be used to judge whether the two measures of coherence give the same ordering or not.

Step 1: To find two states satisfying the following condition:
\begin{equation}\label{panju}
  \mathcal{C}_A(\rho_1)=\mathcal{C}_A(\rho_2)\nLeftrightarrow \mathcal{C}_B(\rho_1)=\mathcal{C}_B(\rho_2).
\end{equation}
The existence of such states signifies
that the two measures do not give the same ordering. Otherwise, to carry out step 2.

Step 2: Sort the concerned states by the measure $\mathcal{C}_A(\rho)$ as an ascending sequence. Then, calculate another measure $\mathcal{C}_B(\rho)$ orderly and get the sequence $\{\mathcal{C}_B(\rho)\}$. If $\{\mathcal{C}_B(\rho)\}$ is also an ascending sequence, it can be declared that the two measures give the same ordering for those states, otherwise, the two measures do not imply the same ordering.
What's more, for continuous states, the sequence $\{\mathcal{C}_B(\rho)\}$ has no extreme point (or the extreme point is starting point or endpoint) and its monotone increasing imply that the two measures give the same ordering.

In Ref. \cite{T. Baumgratz}, a strict framework of coherence has been proposed
to quantify suitable coherence measure,
where the elative entropy of coherence and
the $l_1$ norm of coherence were put forward.
They have been identified as general and easy-calculating measures.

The relative entropy of coherence
is defined \cite{T. Baumgratz}
\begin{equation}\label{}
  \mathcal{C}_{re}(\rho)=\min_{\delta\epsilon\mathcal{I}}S(\rho\|\delta)=S(\rho_{{diag}})-S(\rho),
\end{equation}
where $\rho_{diag}$ comes from $\rho$ by dropping off-diagonal elements, $S(\rho\|\delta)=\Tr(\rho\log\rho-\rho\log\delta)$ is the quantum relative entropy\cite{Nielsen} and $S(\rho)=-\Tr(\rho\log\rho)$ is the von Neumann entropy.
The $l_1$ norm of coherence
is defined \cite{T. Baumgratz}
\begin{equation}\label{}
  \mathcal{C}_{l_1}(\rho)=\min_{\delta\epsilon\mathcal{I}}|\rho-\delta|_{l_1} =\sum_{i\neq j}|\rho_{ij}|,
\end{equation}
where $\rho_{ij}$ are entries of $\rho$.

For two-qubit Bell-diagonal states, the relative entropy of coherence is given by
\begin{eqnarray}\label{coherencesolution}
  \mathcal{C}_{re} =-H(\lambda_{ab})-\sum_{j=1}^2\frac{(1+(-1)^jc_3)}{2}\log_2\frac{(1+(-1)^jc_3)}{4},\n
\end{eqnarray}
where $H(\lambda_{ab})=-\sum_{ab}\lambda_{a,b}\log_2\lambda_{ab}$,
while the $l_1$ norm of coherence is given by
\begin{equation}
  \mathcal{C}_{l_1}=\frac{1}{2}|c_1-c_2|+\frac{1}{2}|c_1+c_2|.
\end{equation}
The results of above equation is summarized in Tab. \ref{Cl1}.

\begin{table}[ht]
\caption{  The solution of $C_{l_1}$ for Bell-diagonal states in four different regions.
}
\begin{ruledtabular}
\begin{tabular}{lcccc}
  \multirow{2}{*}{region} &$c_1-c_2 \geq  0$  &$c_1-c_2 \geq  0$ &$c_1-c_2 \leq0$ &$c_1-c_2 \leq 0$ \\
                          &$c_1+c_2 \geq 0$  & $c_1+c_2 \leq 0 $ &$c_1+c_2 \geq0$ &$c_1+c_2 \leq 0$ \\[3pt]
  \hline
  $\mathcal{C}_{l_1}$ & $c_1$ & $-c_2$ & $c_2$ & $-c_1$ \\[3pt]
\end{tabular}
\end{ruledtabular}
\label{Cl1}
\end{table}

\begin{figure}[ht]
  \centering
  \includegraphics[width=0.49\textwidth]{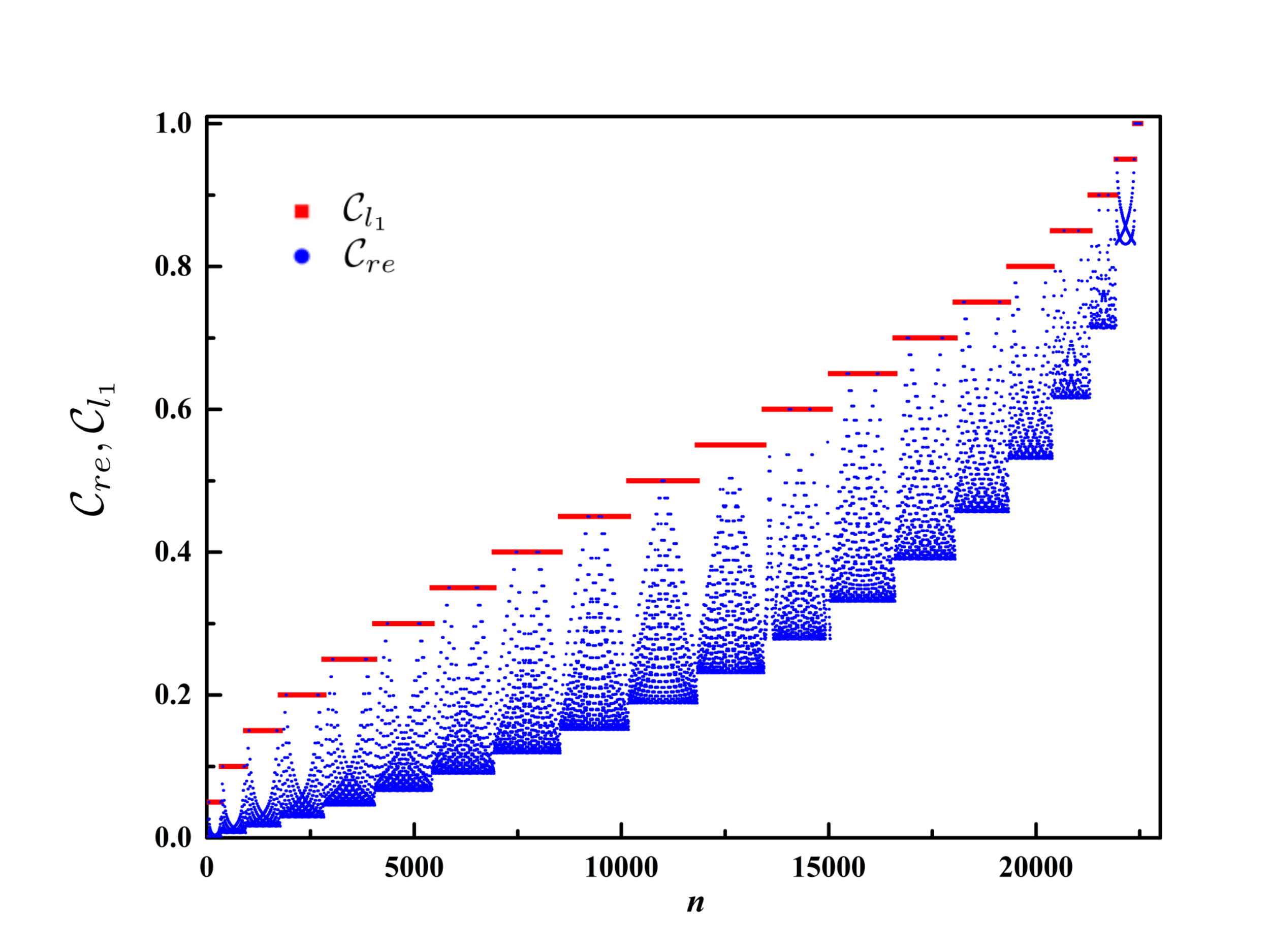}\\
  \caption{The coherence of Bell-diagonal states sorted by $\mathcal{C}_{l_1}$ as a ascending sequence. The red points represent $l_1$ norm of coherence $\mathcal{C}_{l_1}$ for two-qubit Bell-diagonal states, the blue points represent relative entropy of coherence $\mathcal{C}_{re}$, and $n$ is correspond to a state with $n^{th}$ place in comparing the value of $\mathcal{C}_{l_1}$. While $\mathcal{C}_{l_1}$ becomes bigger and bigger but $\mathcal{C}_{re}$ is not monotone increasing.}%
  \label{paixu}
\end{figure}

\begin{figure}[ht]
  \centering
  \includegraphics[width=0.49\textwidth]{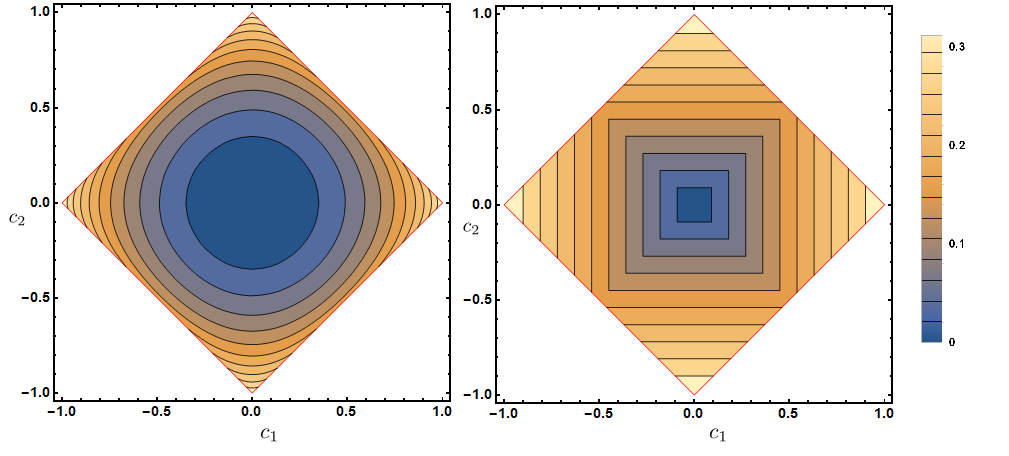}\\
  \caption{The left and right graph are the counterplot of $\mathcal{C}_{re}$ and $\mathcal{C}_{l_1}$ respectively. Those counterplot have no overlap and monotone increasing as one moves outward.}%
  \label{counterplot}
\end{figure}

By using the step 1 of the method mentioned in Sec. \ref{aaa}, the solution of $\mathcal{C}_{l_1}$ tells that $\mathcal{C}_{l_1}$ and $\mathcal{C}_{re}$ do not give the same ordering for the states limited in the lines which are parallel to one of the axes.
We show the step 2 by Fig. \ref{paixu},
in which we can clearly find that $\mathcal{C}_{l_1}$ and $\mathcal{C}_{re}$
do not give the same ordering for all Bell-diagonal states.
One can find either the states for which the ordering
is violated or preserved.
For simplicity, we consider the ordering of coherence for the states limited in the cross section $c_3=0$.
$\mathcal{C}_{re}$ and $\mathcal{C}_{l_1}$ is a monotone decreasing for the states limited to $c_1$ and $c_2$ axes towards the origin of coordinates.
The sets of ray states limited in region $c_2\neq0, c_3=0$ can be parameterized as $c_2=mc_1$ $(m\in R)$, Then, the relative entropy of coherence is given by
\begin{eqnarray}
  \mathcal{C}_{re} &=& \frac{1}{4}[(1-c_1-mc_1)\log_2(1-c_1-mc_1) \n
   &{}& +(1-c_1+mc_1)\log_2(1-c_1+mc_1) \n
   &{}& +(1+c_1-mc_1)\log_2(1+c_1-mc_1) \n
   &{}& +(1+c_1+mc_1)\log_2(1+c_1+mc_1)].
\end{eqnarray}
The first derivative of $\mathcal{C}_{re}$ is
\begin{eqnarray}
  \frac{d\mathcal{C}_{re}}{dc_1} &=& \frac{1}{4}[(-1-m)\log_2(1-c_1-mc_1) \n
   &{}& +(-1+m)\log_2(1-c_1+mc_1) \n
   &{}& +(1-m)\log_2(1+c_1-mc_1) \n
   &{}& +(1+m)\log_2(1+c_1+mc_1)],
\end{eqnarray}
and the second derivative of $\mathcal{C}_{re}$ is
\begin{eqnarray}
  \frac{d^2\mathcal{C}_{re}}{d^2c_1} &=& \frac{1}{4\ln2}[\frac{(-1-m)^2}{1-c_1-mc_1}+\frac{(-1+m)^2}{1-c_1+mc_1} \n
   &{}& +\frac{(1-m)^2}{1+c_1-mc_1}+\frac{(1+m)^2}{1+c_1+mc_1}].
\end{eqnarray}
It is easy to verify that $\frac{d\mathcal{C}_{re}}{dc_1}\mid_{c_1=0}=0 $ and $\frac{d^2\mathcal{C}_{re}}{d^2c_1}>0$. Thus the extremum of $\mathcal{C}_{re}$ is $\mathcal{C}_{re}=0$ if and only if $c_1=0$. It means that $\mathcal{C}_{re}$ and ${\mathcal{C}_{l_1}}$ give the same ordering for the states limited in any one ray from the edge states to the center of tetrahedron.
This can be also verified by the counterplot in Fig. \ref{counterplot}.

\section{the structure of coherence for Bell diagonal states}\label{bbb}


The ordering of coherence measures for Bell-diagonal states do not preserved,
but the relative entropy of coherence and $l_1$ norm of coherence show that coherence of different measures may have an identical structure.
In this section, we consider the structure of coherence which is not affected by the choice of coherence measure.
Monotonicity of coherence means that coherence does not increase through incoherent quantum channels. The trajectory of time evolution under incoherent quantum channels for Bell diagonal states must go through the level surface of coherence unidirectionally and the coherence is monotone decreasing along the trajectory.
The states limited in those trajectory have the same ordering for all coherence measures,
and, thus, we investigate the structure of coherence in terms of incoherent quantum channels.

It is worth-noting that the states limited in $c_3$ axis are incoherent states, and the states at vertices (Bell states) are maximally coherent states.
For the other states, we investigate the structure of coherence by using the fact that incoherent quantum channel do not increase coherence.
Incoherent quantum channels can be characterized by a set of Kraus operators $\{K_j\}$, whose action on the state $\rho$ of the system can be described as
$\Lambda(\rho)=\sum_jK_j\rho K_j^\dag$, which satisfy the constrains $\sum_jK_j^\dag K_j=I$ and $K_j\mathcal{I}K_j^\dag\subset\mathcal{I}$ for all $j$,  where $\mathcal{I}$ is the set of incoherent states.
We consider countless sets of states for which all coherence measure give the same ordering,
and Bell-diagonal states in tetrahedron consist of those sets of states.
We find that depolarizing channel and phase flip channel can be used to explore the structure of coherence.
It is easy to verify that the above two quantum channels are both incoherent, and those two quantum channels act on Bell-diagonal states have no freezing phenomenon.
\subsection{The evolution of states under depolarizing channel}

The depolarizing channel has the elements
\begin{eqnarray}
  K_0 &=& \sqrt{1-3q}I,\quad K_1 = \sqrt{\frac{q}{4}}\sigma_1 ,\n
  K_2 &=& \sqrt{\frac{q}{4}}\sigma_2,\qquad \ K_3 = \sqrt{\frac{q}{4}}\sigma_3 .
\end{eqnarray}
where the parameter $q(t)=1-e^{-\gamma t}$ is the strength of the noise and $\sigma_1$ is the Pauli matrix.
We put depolarizing channel on subsystem $A$, the time evolution of Bell-diagonal states can be expressed as
\begin{eqnarray}
  c_1(t) &=& (1-\frac{q}{2})c_1(0) \\
  c_2(t) &=& (1-\frac{q}{2})c_2(0) \\
  c_3(t) &=& (1-\frac{q}{2})c_3(0)
\end{eqnarray}

The depolarizing channel on subsystem $A$ achieves
the states transformation toward the center of tetrahedron.
The monotonicity of coherence ensures that all coherence measures
give the same ordering and the coherence is monotone decreasing
(except the incoherent states limited in $c_3$ axis) for
the states limited in any ray from the surface of tetrahedron to the center of tetrahedron.

\subsection{The evolution of states under phase flip channel}

The phase flip channel destroys the information contained in the phase relations without an exchange of energy.
The phase flip channel has operation elements
\begin{eqnarray}
  K_{20} &=& \sqrt{1-q(t)/2}I, \n
  K_{21} &=& \sqrt{q(t)/2}\sigma_3,
\end{eqnarray}
where $\sigma_3$ is the Pauli matrix. We put phase flip channel on system $A$ and system $B$ respectively, the time evolution of Bell-diagonal states can be expressed as
\begin{eqnarray}
  c_1(t) &=& c_1(0)e^{-2\gamma t}, \n
  c_2(t) &=& c_2(0)e^{-2\gamma t}, \n
  c_3(t) &\equiv& c_3(0).
\end{eqnarray}

The trajectory of the states under phase flip channel are straight line perpendicular to $c_3$ axis.
The trajectory implies that the coherence is monotone decreasing, while the states close to $c_3$ axis.

\section{the structure of discord for Bell diagonal states}\label{ccc}

We will consider the states limited to any ray from the surface to the center of tetrahedron to explore the structure of discord.
In this section, we focus on
quantum discord and geometric quantum discord.

The quantum mutual information of system $A$ and $B$ is given by
\begin{equation}\label{}
  \mathcal{I}(A:B)=S(A)+S(B)-S(AB).
\end{equation}
And the classical mutual information is given in the following form \cite{henderson}
\begin{equation}\label{a}
  \mathcal{J}_{cl}(A|B)=S(B)-S(B|A),
\end{equation}
where $S(B|A)=\sum_ap_aS(B|a)$ is the conditional entropy \cite{Nielsen}.
If we have a set of positive-operator valued measures (POVM) with elements $E_a=M_a^\dag M_a$ and classical outcome $a$ on subsystem $A$, we can introduce classical correlations of the states $\rho_{AB}$ in analogy to Eq. (\ref{a})

\begin{equation}\label{b}
  \mathcal{J}(B|\{E_a\})=S(B)-S(B|\{E_a\}).
\end{equation}
The quantum discord\cite{H. Ollivier} of a state $\rho_{AB}$ under a measurement $\{E_a\}$ is defined as the difference between total correlations, as
given by the quantum mutual information Eq. (\ref {a}), and the
classical correlations (\ref{b})
\begin{equation}\label{c}
  \mathcal{D}(B|A)\equiv \min_{E_a}\{\mathcal{I}(A:B)-\mathcal{J}(B|\{E_a\})\},
\end{equation}
where $\mathcal{{D}}(B|A)$ is minimized over all measurements.

The geometric quantum discord based Hilbert-Schmidt distance to the states after measure is defined as \cite{Daki}
\begin{equation}\label{d}
  \mathcal{D}^G=\min_{\chi\epsilon\mathcal{C}}\|\rho-\chi\|^2=\min_{\chi\epsilon\mathcal{C}}tr[(\rho-\chi)^2],
\end{equation}
where $\mathcal{C}$ is the set of classical-quantum states.

For two-qubit Bell-diagonal states, the quantum discord is given by \cite{luo}
 \begin{eqnarray}\label{discordsolution}
  \mathcal{D}(\rho_{AB})=-H(\lambda_{ab})-\sum_{j=1}^2\frac{(1+(-1)^jc)}{2}\log_2\frac{(1+(-1)^jc)}{4},\n
\end{eqnarray}
where $c=\max\{|c_1|,|c_2|,|c_3|\}$.
And the geometric quantum discord is given by \cite{Daki}
\begin{equation}\label{geometeric discord}
  \mathcal{D}^G(\rho_{AB})=\frac{1}{4}(c_1^2+c_2^2+c_3^2-c^2).
\end{equation}

We discuss the Bell-diagonal states in three cases to investigate the structure of quantum discord. For the easy of expression, we define the states limited to any one ray toward the center of tetrahedron as any one set of ray states.

\textbf{Case 1}:  Any one set of ray states limited in coordinate axis.

It is easy to verify that $\mathcal{D}(\rho_{AB})=\mathcal{D}^G(\rho_{AB})=0$ for the sets of ray states limited in coordinate axis. As for this case quantum discord and quantum geometric discord are unified.

\textbf{Case 2}: Any one set of ray states limited in region $c_2\neq0,c_3=0$.

\begin{figure}[ht]
  \centering
  \includegraphics[width=0.49\textwidth]{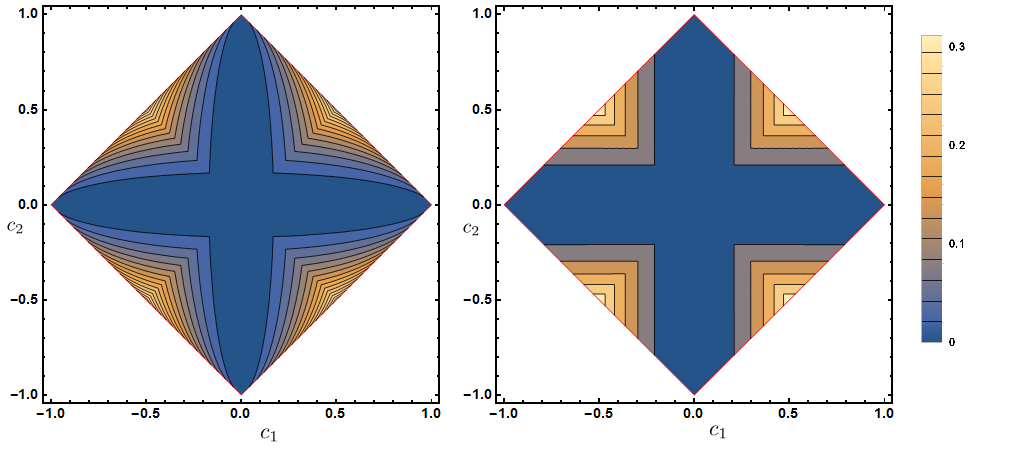}\\
  \caption{The left and right graph are the counterplot of quantum discord and geometric quantum discord respectively. Those counterplot also have no overlap and monotone increasing as one moves outward.}%
  \label{counterplot2}
\end{figure}

Any set of ray states limited in region $c_2\neq0,c_3=0$, can be parameterized as $c_1=mc_2$, then geometric quantum discord (\ref{geometeric discord}) is
\begin{eqnarray}
  \mathcal{D}^G(\rho_{AB}) = \left\{\begin{array}{ll}
\frac{1}{4}c_2^2 & c=c_1\\
\frac{1}{4}m^2c_2^2 & c=c_2\\
\frac{1}{4}(m^2+1)c_2^2  & c=c_3.
\end{array}\right.
\end{eqnarray}
It is obvious that geometric quantum discord and quantum discord are both monotone decreasing for any set of states toward to the center of tetrahedron in this case. This can be directly verified by the contourplot in Fig. \ref{counterplot2}.

\textbf{Case 3}: The rest generic sets of ray states limited in Bell-diagonal states.

\begin{figure}[ht]
  \centering
  \includegraphics[width=0.49\textwidth]{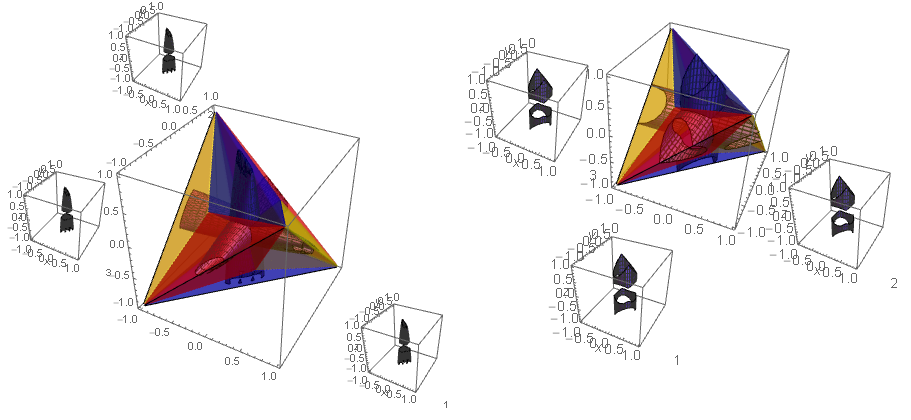}\\
  \caption{The left and right graph are the level surface of $\mathcal{D}=0.03$ and $\mathcal{D}^G=0.03$ respectively.}%
  \label{discordlevelsurface}
\end{figure}

The rays limited in $\{c_i\neq0, i=1,2,3\}$ are parameterized as $c_1=ac_3, c_2=bc_3,a\in R,b\in R $, as a set of ray states, then geometric quantum discord (\ref{geometeric discord}) is
\begin{eqnarray}
  \mathcal{D}^G(\rho_{AB}) = \left\{\begin{array}{ll}
\frac{1}{4}(b^2+1)c_2^2 & c=c_1\\
\frac{1}{4}(a^2+1)c_3^2 & c=c_2\\
\frac{1}{4}(a^2+b^2)c_3^2  & c=c_3.
\end{array}\right.
\end{eqnarray}
The result shows that geometric quantum discord is monotone decreasing for any one set of ray states.
It is easy to verify that quantum discord is also  monotone decreasing for any one set of ray states.
In Fig \ref{discordlevelsurface}, we plot the level surface of quantum discord and geometric quantum discord.
Combining with Fig. \ref{counterplot2}, one can see that any one ray towards the center of tetrahedron go through the level surface of discord and
the discord is monotone decreasing for any set of ray states.


\section{Conclusion}\label{eee}

We propose a scenario to judge whether the two measures of coherence give the same ordering or not.
Using our scenario we verify that two coherence measures, the relative entropy of coherence and $l_1$ norm of coherence, do not give the same ordering for all Bell-diagonal states.

Considering the monotonicity of coherence, we show that all coherence measures give the same ordering for any one set of ray states and for the states limited in any one line which is perpendicular to $c_3$ axis.
Notably, geometric quantum discord and quantum discord give the same ordering for any one set of ray states.

We picture the structure of coherence and quantum discord for Bell-diagonal states as follows.
The coherence is zero for the states limited in $c_3$ axis. 
As the states close to the center of tetrahedron and $c_3$ axis, the coherence is monotone decreasing asymptotically.
The level surface of coherence have no overlap, and can not through $c_3$ axis.
The discord is zero for the states limited in three axes. 
As the states close to the center of tetrahedron and three axes, the discord is monotone decreasing asymptotically.
The level surface of coherence have no overlap, and can not through three axes.

Our work provides a complete and unified structure of coherence and quantum discord with different measures for Bell-diagonal states.
The behavior of quantum resources under quantum channel has been widely studied for past few years.
The pictorial structure of coherence and discord tells us the behavior of resources along the trajectory directly.
The trajectory of time evolution of quantum states, which only depends on initial states and quantum channel, is visualized to understand the behavior of quantum resources.
We will deeply study the behavior of resources by the guidance of our results in the future.

\section{Acknowledgements}
We thank Hai-Long Shi and Guo-Guo Xin, for their valuable discussions. This work was
supported by the NSFC (Grant No.11375141 and No.11425522), the Special Research Funds of Shaanxi Province Department of Education (No.203010005), Northwest University Scientific Research Funds (No.338020004) and the Double First-Class University Construction Project of Northwest University.


\begin{thebibliography}{99}


\bibitem{R. Horodecki}{R. Horodecki, P. Horodecki, M. Horodecki, and K. Horodecki, Rev. Mod. Phys. \textbf{81}, 865 (2009).}
\bibitem{C. H. Bennett}{C. H. Bennett, Phys. Rev. Lett. \textbf{68}, 3121 (1992).}
\bibitem{C. H. Bennett2}{C. H. Bennett and S. J. Wiesner, Phys. Rev. Lett. \textbf{69}, 2881 (1992).}
\bibitem{C. H. Bennett3}{C. H. Bennett, G. Brassard, C. Cr\'{e}peau, R. Jozsa, A. Peres, and W. K. Wootters, Phys. Rev. Lett. \textbf{70}, 1895 (1993).}
\bibitem{H. Ollivier}{H. Ollivier and W. H. Zurek, Phys. Rev. Lett. \textbf{88}, 017901 (2001).}

\bibitem{A. Streltsovdiscord}{A. Streltsov, G. Adesso, and M. B Plenio, Rev. Mod. Phys. \textbf{84}, 1655 (2012).}

\bibitem{M. Piani}{M. Piani, P. Horodecki, and R. Horodecki, Phys. Rev. Lett. \textbf{100}, 090502 (2008).}
\bibitem{A. Ferraro}{A. Ferraro, L. Aolita, D. Cavalcanti, F. M. Cucchietti, and A. A\'{c}n, Phys. Rev. A \textbf{81}, 052318 (2010).}
\bibitem{F. Altintas}{F. Altintas, Opt. Commun. \textbf{283}, 5264 (2010).}
\bibitem{Y. Yeo}{Y. Yeo, J.-H. An, and C. H. Oh, Phys. Rev. A \textbf{82}, 032340 (2010).}
\bibitem{B. Bellomo}{B. Bellomo, R. L. Franco, and G. Compagno, Phys. Rev. A \textbf{85}, 024302 (2012).}
\bibitem{Y. Li}{Y. Li, B. Luo, and H. Guo, Phys. Rev. A \textbf{84}, 012316 (2011).}


\bibitem{T. Baumgratz}{T. Baumgratz, M. Cramer, and M. B. Plenio, Phys. Rev. Lett. \textbf{113}, 140401 (2014).}



\bibitem{A. Winter}{A. Winter and D. Yang, Phys. Rev. Lett. \textbf{116}, 120404 (2016).}
\bibitem{Fanheng1}{K. F.  Bu, H. Fan, A. K. Pati, and J. Wu, arXiv: 1610.09646 (2016).}
\bibitem{A. Streltsov}{A. Streltsov, Gerardo Adesso, and Martin B. Plenio, arXiv: 1609.02439 (2017).}
\bibitem{shi}{H. L. Shi, S. Y. Liu,  X. H. Wang,  W. L. Yang and H. F, Phys. Rev. A \textbf{95}.032307}



\bibitem{Horodecki}{R. Horodecki, and M. Horodecki, Phys. Rev. A \textbf{54}, 1838 (1996).}
\bibitem{T. R. Bromley}{T. R. Bromley, M. Cianciaruso, and G. Adesso (2015), Phys. Rev. Lett. \textbf{114}, 210401.}
\bibitem{L. Mazzola}{L. Mazzola, J. Piilo, and S. Maniscalco, Phys. Rev. Lett. \textbf{104}, 200401 (2010).}

\bibitem{Wootters}{W. K. Wootters, Phys. Rev. Lett. \textbf{80}, 2245 (1998).}
\bibitem{Matthias D. Lang}{M. D. Lang and C. M. Caves, Phys. Rev. Lett. \textbf{105}, 150501}

\bibitem{D. Girolami}{D. Girolami, Phys. Rev. Lett. \textbf{113}, 170401 (2014).}
\bibitem{A. Streltsov1}{A. Streltsov, U. Singh, H. S. Dhar, M. N. Bera and G. Adesso, Phys. Rev. Lett. \textbf{115}, 020403 (2015).}
\bibitem{L.-H. Shao}{ L.H. Shao, Z. Xi, H. Fan and Y. Li, Phys. Rev. A \textbf{91}, 042120 (2015).}
\bibitem{X. Yuan}{X. Yuan, H. Zhou, Z. Cao, and X. Ma, Phys. Rev. A \textbf{92}, 022124 (2015).}

\bibitem{U. Singh}{U. Singh, M. N. Bera, H. S. Dhar, and A. K. Pati, Phys. Rev. A \textbf{91}, 052115 (2015).}
\bibitem{S. Cheng}{S. Cheng and M. J. W. Hall, Phys. Rev. A \textbf{92}, 042101 (2015).}
\bibitem{M. N. Bera}{M. N. Bera, T. Qureshi, M. A. Siddiqui, and A. K. Pati, Phys. Rev. A \textbf{92}, 012118 (2015).}
\bibitem{A. Mani}{A. Mani and V. Karimipour, Phys. Rev. A \textbf{92}, 032331 (2015).}
\bibitem{Daki}{B. Daki\'{c}, V. Vedral, and  C. Brukner, Phys. Rev. Lett. \textbf{105}, 190502 (2010).}






\bibitem{Bell state} { U. Fano, Rev. Mod. Phys. \textbf{55}, 855 (1983).}
\bibitem{Nielsen}{M. A. Nielsen and I. Chuang, Quantum Computation and Quantum Information (Cambridge University Press, Cambridge, England) (2000).}

\bibitem{henderson}{L. Henderson and V. Vedral, J. Phys. A \textbf{34}, 6899 (2001).}


\bibitem{luo}{S. L. Luo, Phys. Rev. A \textbf{77}, 042303 (2008).}





















\end{thebibliography}
\end{document}